\documentstyle[preprint,aps,pra]{revtex}
\tightenlines
\begin{document}

\draft

\title{Oscillator neural network model with distributed 
native frequencies}

\author{Michiko Yamana, Masatoshi Shiino and Masahiko Yoshioka}
\address{Department of Physics, Tokyo Institute of Technology, Meguro,
Tokyo, Japan} 

\date{\today}

\maketitle

\begin{abstract}
We study associative memory of an oscillator neural network with 
distributed native frequencies. 
The model is based on the use of the Hebb learning rule with random patterns
($\xi_i^{\mu}=\pm 1$),
and the distribution function of native frequencies is assumed to be symmetric
with respect to its average.
Although the system with an extensive number of stored patterns
is not allowed to get entirely synchronized, 
long time behaviors of the macroscopic order parameters
describing partial synchronization phenomena can be obtained
by discarding the contribution from the desynchronized part of the system.
The oscillator network is shown to work as associative memory
accompanied by synchronized oscillations.
A phase diagram representing properties of memory retrieval 
is presented in terms of the parameters characterizing the
native frequency distribution.
Our analytical calculations based on the self-consistent signal-to-noise analysis
are shown to be in excellent agreement with numerical simulations, confirming 
the validity of our theoretical treatment.

\end{abstract}

\bigskip

\section{Introduction}
Most of the theoretical models for associative memory of
neural networks as typified by the Hopfield model~\cite{hopf} have been based
on the idea of rate coding, 
which assumes that information is coded in the firing rate of a neuron at a 
particular time.
On the other hand, some experimental results on the visual cortex 
have been suggesting the possibility of another scheme of information coding,
that is, the concept of temporal coding
which assumes that information is coded in the relative timing of the firing 
pulses~\cite{singer}.

Stimulated by these experimental findings, studies
of the temporal coding are becoming an active area of 
theoretical brain researches.
Recently, there have been an increasing number of papers reporting 
neurophysiological 
experimental findings of
synchronization phenomena of neurons in animal's brain~\cite{monkey}.
Synchronized firings of neurons can be considered to play a key role in certain
types of information processing such as the binding problem~\cite{bind}.

In order to get insights into basic properties of the scheme of the temporal
coding it will be necessary to study neural network models based on neurons
that can be expressed by such nonlinear oscillators as limit-cycle type
and integrate-and-fire type oscillators.
Indeed, a network of integrate-and-fire neurons, 
where one can be directly concerned 
with the spike timing of neurons, is one of such models and was previously
studied~\cite{YS}
to confirm its functioning as associative memory accompanied by synchronized 
firings.

The simplest theoretical model of synchronization of coupled oscillators
will be the case of
a system of phase oscillators of Kuramoto et al.~\cite{kura,daido,saka-kura},
which is derived
under the assumption of weak coupling limit for general types of limit-cycle
oscillators.
In an associative memory model
based on such a system with the Hebb type learning
rule, the distribution of native frequencies will become crucial for
the behavior of
synchronization or desynchronization of the system~\cite{kura,shii,aoyagi,vicente}.
If all of the oscillators have an identical native frequency and the couplings are 
symmetric, the network can be reduced, by an appropriate coordinate transformation,
to a system that has a Lyapunov function ensuring stability of 
its equilibrium state(s),
which implies that the network eventually
settles into a synchronized state.
On the other hand, in the case when native frequencies are distributed,
Lyapunov functions in general no longer exist except for a particular 
case~\cite{vanhemm}
and the network becomes essentially of
dynamic nature;
one may expect partial synchronization or more complex oscillatory behavior.

In the present paper, we report results of our detailed analysis
of associative memory of an oscillator neural network model
with distributed native frequencies.
Whereas oscillator networks with a Lyapunov function were studied previously
by several authors~\cite{cook,aoni}, 
satisfactory analysis of the case without a Lyapunov function
has been far less conducted~\cite{AKO}.

\section{Analysis of the model}
The oscillator network model we consider is described by the following phase
evolution equations~\cite{kura}:
\begin{equation}
\frac{d\phi_i}{dt}=\omega_i-\sum_j J_{ij}\sin(\phi_i-\phi_j), ~~~~~~~(i=1,\cdots,N) 
\label{phase-eq}
\end{equation}
where $N$ denotes the total number of neurons, and 
$\phi_i$ and $\omega_i$ respectively
denote the phase and the native frequency of $i$-th neuron.
We assume the strength of interactions to be given according to 
the Hebb learning rule,
\begin{equation}
J_{ij}=\frac{1}{N} \sum_{\mu} \xi_i^{\mu} \xi_j^{\mu},
\end{equation}
where $\{\xi_i^{\mu}\ \}$ ($\mu=1\cdots p$) represents the $\mu$-th
stored patterns.
We consider the case when 
the number of patterns is extensive, $p=\alpha N$.
In order to elucidate the effect of the distribution of native frequencies
on the behavior of synchronization we want to make the model as simple as
possible, and then we assume $\xi_i^{\mu}$ to take values $\pm 1$ 
rather than continuous values 
with $\xi_i^{\mu} = \exp(i \theta_i^{\mu})$,
($\theta_i^{\mu}=[0: 2\pi]$), which would be more appropriate for the 
study of temporal coding itself.

Furthermore, 
we deal with a simple case where the native frequency distribution $p(\omega)$ 
is discrete and symmetric with respect to the central frequency $\omega_0$:
$p(\omega)=\sum_{k=-L}^{L} C_k \delta(\omega-\omega_k)$,
$\omega_k+\omega_{-k}=2\omega_0,~C_k=C_{-k}\ge 0,~\sum_{k=-L}^{L} C_k=1$.   
It is noted that without loss of generality one can set $\omega_0=0$ as a
result of rotational symmetry of the phase evolution equation (\ref{phase-eq}).
If the native frequencies are the same,
i.e., $\omega_i\equiv\Omega$ independent of $i$,
all of the oscillators can get synchronized with
$\phi_i(t)=\phi_i^0 +\Omega t$.
In the case of distributed native frequencies, 
there remain, in general, a group of desynchronized oscillators.
Then the total system cannot settle into entirely equilibrium states.

Assuming that influences exerted by a group of desynchronized neurons 
on the macroscopic behavior of the total system can be neglected,
we use the self-consistent signal-to-noise analysis (SCSNA)~\cite{fuka-shii}
to analytically obtain the macroscopic order parameter equations 
for the time-stationary states of synchronized oscillations of the 
network accompanying memory retrieval.

Introducing the complex variable $x_k \equiv \exp(i \phi_k)$
to express the state of $k$-th neuron,
we formally obtain the fixed point equations by putting $\dot{x}_i =0$;
\begin{equation}
x_i=\frac{i \omega_i + \sqrt{ |h_i|^2-\omega_i^2 }}
{{\tilde h_i}}, ~~~~~|h_i| > \omega_i,
\label{equilibrium}
\end{equation}
where $h_i \equiv \sum_j J_{ij} x_j$ is the local field
and ${\tilde h_i}$ denotes its complex conjugate.
Defining the overlap
\begin{equation}
m^{\mu}=\frac{1}{N}\sum_i\xi_i^{\mu} x_i,
\end{equation}
one can rewrite the local field as 
$h_i=\sum_{\mu}\xi_i^{\mu}m^{\mu}$.

Within the frame work of the SCSNA,
considering the retrieval solutions of Eq.~(\ref{equilibrium})
for which
$m^{1}=m\sim O(1)$ and $m^{\mu}\sim O(1/\sqrt{N})(\mu\neq1)$,
we assume the local field to be described as~\cite{fuka-shii}
\begin{equation}
h_i=\xi_i^1 m+z_i + \Gamma_1 x_i +\Gamma_2 {\tilde x_i}.
\label{local}
\end{equation}
Here the first term involving $m$ is the signal part, while the remaining terms
represent noise part
involving complex Gaussian noise $z_i=u_i+iv_i$ ($u_i,~v_i$, real)
together with the effective self-coupling terms proportional 
to $x_i$ and its complex conjugate $\tilde x_i$.
We note that Eq.~(\ref{equilibrium}) yield no solutions 
if $|h_i| < \omega_i$, which does not ensure
$|x_i| = 1$.
This means that neurons with $|h_i| < \omega_i$  cannot take part in 
synchronized motions exhibited by neurons with $|h_i| > \omega_i$.
Although the desynchronized neurons, each of 
which is expected to oscillate with a 
certain modified frequency, will make the local fields 
time dependent quantities, their effect can be expected to 
cancel out to good approximation provided taking the 
time average is considered.
In further analysis, we discard the 
contribution from the desynchronized neurons by setting $x_i=0$
for $|h_i| < \omega_i$.

Following the standard procedure of the SCSNA, we obtain the 
self-consistent equations for macroscopic variables in the limit
$N \to \infty$.
Making use of the rotational symmetry of the phase evolution equation
Eq.~(\ref{phase-eq}) to choose a gauge such that overlap $m$ becomes
real i.e.
$\left[\int {\rm D}u {\rm D}v~ \xi^1 \sin \phi \right]_{\{\omega,~\xi^1\}} =0$,
we have
\begin{mathletters}
\begin{eqnarray}
&&m = \frac{1}{N}\sum_i \xi_i^1 x_i(\xi_i^1,~\omega_i,~z_i,
{\tilde z_i})= 
\left[\int {\rm D}u {\rm D}v~\xi^1 \cos \phi \right]_{\{\omega,~\xi^1 \}}, \\
&&q = \frac{1}{N}\sum_i ({\rm Re}[x_i])^2= \left[\int {\rm D}u {\rm D}v
 (\cos \phi)^2
\right]_{\{\omega, \xi^1 \}}, \\
&&U_1 = \frac{1}{N}\sum_i \frac{\partial x_i}{\partial z_i}
=\left[ \int \frac{{\rm D}u{\rm D}v}
{2 \left\{ (\xi^1 m +u)\cos\phi+v\sin\phi
\right\}+4\Gamma_2\cos2\phi}\right]_{\{\omega,~\xi^1 \}}, \\
&&U_2= \frac{1}{N}\sum_i \frac{\partial x_i}
{\partial{\tilde z_i}}
=\left[ \int \frac{-\cos2\phi {\rm D}u{\rm D}v}
{2 \left\{ (\xi^1 m +u)\cos\phi+v \sin\phi \right\}
 +4\Gamma_2\cos2\phi } \right]_{\{\omega,~\xi^1 \}}, \\
&&\Gamma_1=\frac{\alpha (1-U_1)}
{ (1-U_1)^2-U_2^2}, \\
&&\Gamma_2=\frac{\alpha U_2}{ (1-U_1)^2-U_2^2}, \\
\label{gammp}
&&Q_1 = \frac{\alpha q}{ (1-U_1-U_2)^2}, \\
&&Q_2 = \frac{\alpha (1-q)}{ (1-U_1+U_2)^2}, \\
&&{\rm D}u{\rm D}v = \frac{{\rm d}u{\rm d}v}
{2\pi \sqrt{Q_1Q_2}}\exp \left[-\frac{1}{2}
\left( \frac{u^2}{Q_1}+\frac{v^2}{Q_2} \right) \right],
\end{eqnarray}
\end{mathletters}
with $\phi(\xi^1,\omega,u,v)$ being implicitly determined by 
\begin{equation}
f(\phi)\equiv-\omega+(\xi^1 m+u) \sin \phi
-v \cos \phi+\Gamma_2 \sin 2\phi=0.
\label{cond2}
\end{equation}
In the above equations,
$[~~]_{\{\omega,~\xi^1 \}}$ means taking the average over the distribution 
$p(\omega)$ and the pattern $\{ \xi^1 \}$, and
the gaussian integration ${\rm D}u{\rm D}v$ is to be performed over the
noise $u,~v$ satisfying the condition $|h_i| > \omega_i$.
It is noted that in performing the gaussian integration in Eqs.~(6a)-(6i),
one has to take into account the Maxwell rule to pick up the relevant
solution $\phi_i$, when Eq.~(\ref{cond2})
admits multi-solutions owing 
to the presence of the self-coupling term ($\Gamma_2$ term).
Unlike $\Gamma_2$ term, $\Gamma_1$ term has no contribution
to the equilibrium fixed-point equation (7).

In what follows, for the sake of simplicity we deal mainly with the case 
\begin{equation}
p(\omega)=C_0 \delta(\omega)+
\frac{1-C_0}{2} \delta(\omega-\omega_1)+\frac{1-C_0}{2}\delta(\omega+\omega_1),
~~~~~
(L=1,~ \omega_0=0). 
\end{equation}
A characteristic feature of the distribution is that the presence of
oscillators with central frequency $\omega_0$ is allowed with a finite
fraction $C_0$ and the effect of desynchronized part can be described 
in terms of $\omega_1$ and $C_0$ alone.

Setting $\omega_1=0$ recovers the case that allows an energy function
that is bounded from below,
$E[\{\phi_i\}]\equiv -\frac{1}{2}\sum_{ij}J_{ij}\cos(\phi_i-\phi_j)$.

Then all of the oscillators get synchronized for large times with the equilibrium 
configuration $\{\phi_i\}$
determined by Eq.~(\ref{equilibrium}) or Eq.~(\ref{cond2}) together with the 
Maxwell rule, which is explained in Fig.~1.
Analysis based on the SCSNA
of such networks as having Lyapunov functions is
on the same level of approximations as the replica symmetric 
calculations~\cite{fuka-shii,max2,kuhn}.
Figure 2 represents the dependence of $m$ on the loading rate $\alpha$
computed from the SCSNA equations (6) and (\ref{cond2}).
We see that the storage capacity is given by 
$\alpha_c=0.0395$ with $m_c=0.68$.
The present value of the storage capacity slightly
differs from the result reported
previously~\cite{aoni}, which was obtained by an inappropriate treatment of the 
Maxwell rule~\cite{maxwel}.
We note that the Maxwell rule becomes necessary only for $\alpha$
in the neighborhood of $\alpha_c$.
Under the assumption of random patterns with $\xi_i^{\mu} =\pm1$, 
the magnitude of $\Gamma_2$ plays a crucial
role in determining the $\alpha_c$, as can be seen in Fig.~2.
This should be compared with the case of $\xi_i^{\mu}= \exp(i \theta_i^{\mu})$
($\theta_i^{\mu}=[0:2\pi]$), where the order parameter equations involve no
$\Gamma_2$ term as in the case of Cook's model~\cite{cook}
(see also recent work in Ref.~\cite{AKO}).
We note here that when discrete patterns with $Q$-state variables 
$\theta_i^{\mu}=2\pi n_i^{\mu}/Q,~(n_i^{\mu}=0,~\cdots,~Q-1)$ are considered,
the $\Gamma_2$ term does not appear in the order parameter equations, if $Q\geq 3$.

The effect of distribution of native frequencies on the behavior of
memory retrieval is summarized in Fig.~3, where overlap $m$ from the SCSNA
equations is plotted as a function of $\alpha$ and $\omega_1$ for $C_0=0.7$.
There appear two distinct retrieval regimes separated by a valley or gap located 
around a region with
an intermediate value of $\omega_1$ in the $m$-$\alpha$-$\omega_1$ space.
In the regime with small $\omega_1$, most of the oscillators undergo synchronized
motions.
The storage capacities $\alpha_c$ corresponding to the edge of the $m$ surface
are observed to decrease, as $\omega_1$ is increased, to attain a certain
minimum value, where a crossover to the other regime occurs.
In the regime with large $\omega_1$, oscillators with $\omega_1$ oscillate with
their own frequencies
that are modified from the original $\omega_1$
as a result of entrainment phenomenon.
Such a behavior can easily be understood by considering the system in the limit 
$\omega_1 \to \infty$.
In the large $\omega_1$ limit, while most of the oscillators with 
$\omega_i=\omega_0$ get
synchronized, the 
oscillators with $\omega_i=\pm \omega_1$ cannot get synchronized and 
they oscillate with their own frequency $\pm \omega_1$
independently of the synchronized neurons with $\omega_i=\omega_0$.
It is clearly seen 
in this case that the desynchronized neurons does not contribute to the
time-averaged local fields acting on the synchronized neurons.
Then the system can be viewed as a diluted system with only a fraction $C_0$
of neurons participating in memory retrieval.
It is interesting to note that the storage capacity $\alpha_c$ 
in the large $\omega_1$
region increases as $\omega_1$ is increased.
The crossover between the two regimes or the gap in the $m(\omega_1)$
curve with fixed $\alpha$ can be more clearly seen in Fig.~4, which display the 
$\omega_1$ dependence of overlap $m$ obtained from the SCSNA 
together with 
the result of numerical simulations ($N=2000$) for a fixed value of $\alpha$.

As the fraction $C_0$ of the oscillators with the central 
frequency $\omega_0$ is varied,
there occur three types of behavior of the $m(\omega_1)$ curve
showing the cross over between the two regimes.
In Fig.~5, we show the $\omega_1$ dependence of the overlap $m$ obtained
for $C_0=0.5, 0.7$ and $0.9$ with $\alpha=0.02$.
While for $C_0$ large one sees a continuous crossover 
between the small $\omega_1$ regime and the large $\omega_1$ regime,
for $C_0$ small the large $\omega_1$ regime disappears.
For only intermediate values of $C_0$, one observes the gap in the $m(\omega_1)$
curve to appear.

The occurrence of such two retrieval regimes is not restricted to the 
simplest case of $L=1$, but is also observed for 
more general case with $L \ge 2$.
When a sufficiently large number of oscillators have native frequencies near
the central one $\omega_0$,
the system can behave the same way as that of $L=1$ with 
$C_0 \neq 0$,
as is shown Fig.~6, 
where the $\omega_2$ dependence of the overlap $m$ obtained by numerical
simulations is displayed for the native frequency distribution with
$L=2$ and $C_0=0$:

\begin{equation}
p(\omega)=C_1\delta(\omega-\omega_1)+C_1\delta(\omega+\omega_1)
+\frac{1-2 C_1}{2}\delta(\omega-\omega_2)
+\frac{1-2 C_1}{2}\delta(\omega+\omega_2).
\label{L=2}
\end{equation}
Indeed we see a gap separating the small $\omega_2$ regime from the
large $\omega_2$ one.
Although there are no oscillators with $\omega_0=0$ because of $C_0=0$,
most of the oscillators get synchronized for the small $\omega_2$ regime.

\section{Summary and Discussions}
We have studied the behavior of the oscillator neural network system
with distributed native frequencies.
To date analysis of associative memory models has been mostly confined to 
the case of networks with an energy or Lyapunov function that allows
use of replica symmetric calculations of statistical mechanics.
In the present work,
making use of the method of the SCSNA that is free from 
the energy concept~\cite{kura,YS2}, 
we have succeeded in analyzing properties of 
{\it the associative memory accompanied 
by synchronized oscillations} in the prototype oscillator network 
that has no Lyapunov functions except for the case when $p(\omega)=\delta
(\omega-\omega_0)$. 
{\it We have shown that the oscillator network can work
as associative memory based on temporal coding of simple type
even in the presence of distribution of native frequencies.}
Our approach has taken advantage of the fact that such temporal 
attractors as limit-cycles of certain types of dynamical systems can be 
reduced to a fixed-point type attractors as a result of the system's 
symmetry property.  

The distribution of native frequencies does not allow the present coupled
oscillator system to settle into an entirely synchronized state but into a
partially synchronized one.
The contributions from desynchronized neurons to the macroscopic behavior of the
system have, however, been found to be almost negligible.
Thereby, the partially synchronized state of the system has turned out to be
almost determined by the long time behavior of the group of synchronized neurons,
which can be described by fixed-point type attractors giving rise to retrieval
states. 
In other words, {\it memory retrieval is achieved by synchronization of oscillatory
motions of neurons.}

Under the assumed type of native frequency distribution we
have found that {\it the partial synchronization is classified into a high
degree of synchronization that occurs for small $\omega_1$ with overlap $m$
large and a low degree of synchronization that occurs for large $\omega_1$ with
$m$ small.}

Finally we note that oscillator neural networks are considered to have 
advantage over fixed-point type neural networks 
in several respects.
First, oscillator neural networks exhibit 
the ability to easily and efficiently discriminate 
a successful retrieval from unsuccessful one, 
because the settling into a retrieval state of oscillator neural networks
implies the appearance of oscillations with an appreciable amplitude and
the central frequency $\omega_0$ 
in the overlap, and hence, the local fields of neurons.

Second, by utilizing phase as well as amplitude as dynamical variables 
representing output of a neuron~\cite{aoyagi2} 
it becomes possible for neural information to be processed 
in terms of spatio-temporal patterns of neuronal firings.  
In particular, information on time domain is available by employing 
the scheme of temporal coding~\cite{vicente,cook,AKO,aoyagi3}, 
where assuming uniformly distributed random numbers $\theta_i^{\mu}$ 
on $[0, 2\pi)$ for components of the memory patterns, 
the phase difference $\phi_i-\phi_j$ between the two oscillators 
$i$ and $j$ eventually settles into the difference
$\theta_i^{\mu}-\theta_j^{\mu}$ of the memory pattern $\mu$.  
Our present model setting based on the use of binary patterns with 
$\theta_i^{\mu}=0,~\pi$~\cite{aoni}
presents a special as well as simple case of the above temporal coding, 
where the pattern of synchronization is either  of in phase or out of phase.  
Such a case may be related to the problem of segmentation of an object 
from its background~\cite{mals1,mals2,global}.  

More generally, pattern segmentation~\cite{models}
seems to be one of the unique features 
of oscillator neural networks as has been studied by Wang et al.~\cite{wang}, 
who pointed out that use of limit cycles as attractor states of associative 
memory  facilitates switching between one pattern and another on the time domain.  

While studying the important issue of the functionality of pattern segmentation
provided by oscillator neural networks will require taking into account 
some specific ingredients such as
the sparseness of the memory patterns and appropriate inhibitory couplings, 
our simple model will have wide applicability in exploring the computational
ability or relevance exhibited by oscillator neural networks
from the viewpoint of analytical studies.
Extending the assumed symmetric native frequency distribution 
to more general cases of asymmetric one 
as well as continuous ones is straightforward.
Results of such issues together with details of the present work
will be reported elsewhere.

\newpage
\centerline{\bf FIGURE CAPTIONS}
\begin{itemize}
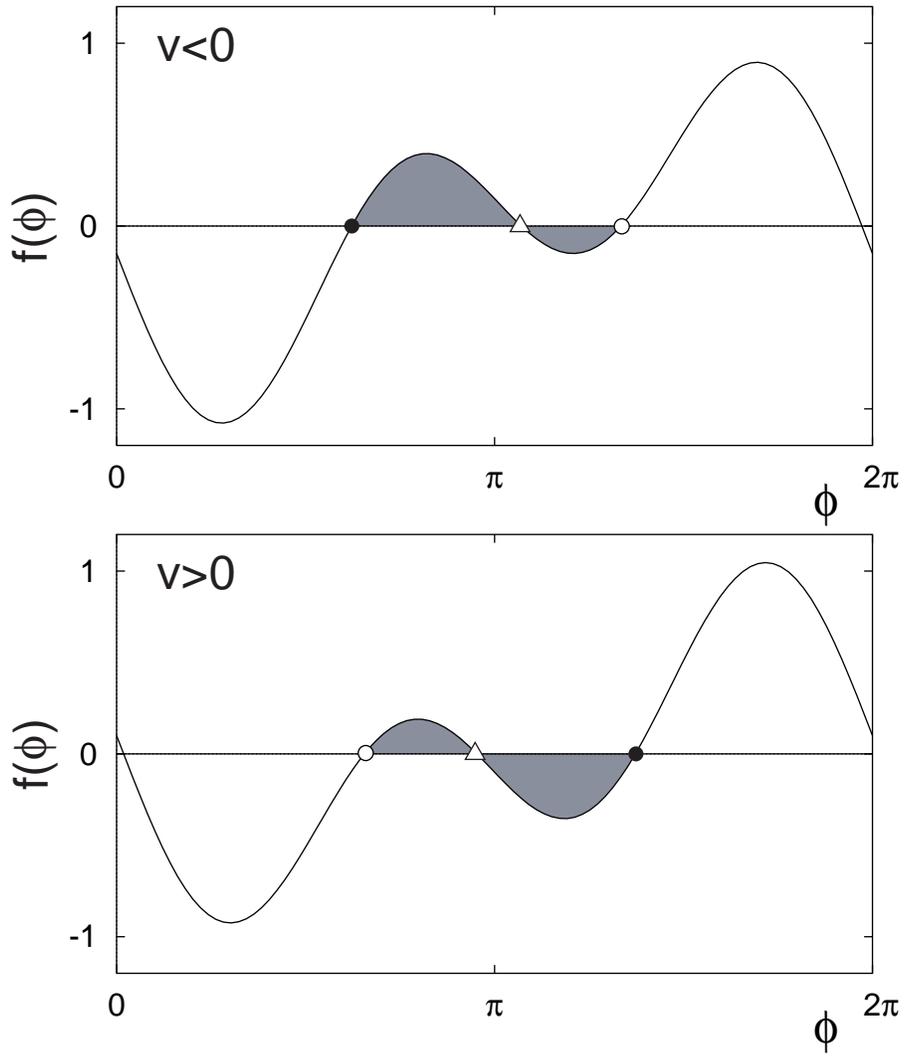
\item[FIG.1:]
Maxwell rule used in the SCSNA for picking up the relevant solution for output
$x_i$ as a function of $u$ and $v$ among multi-solutions 
to Eq.~(\ref{cond2}). In a graphical representation of solving Eq.~(\ref{cond2})
for which three solutions (given by the intersections of the two curves)
appear as $u$ or $v$ is varied, the relevant solution is chosen as the external
intersection that delimits the larger area enclosed between the two curves.
Such solution is marked by the filled circle.
The Maxwell rule ensures the condition of the free energy minimum~\cite{max2}
in a system
with a Lyapunov function as in the case of the liquid-vapor phase transition.

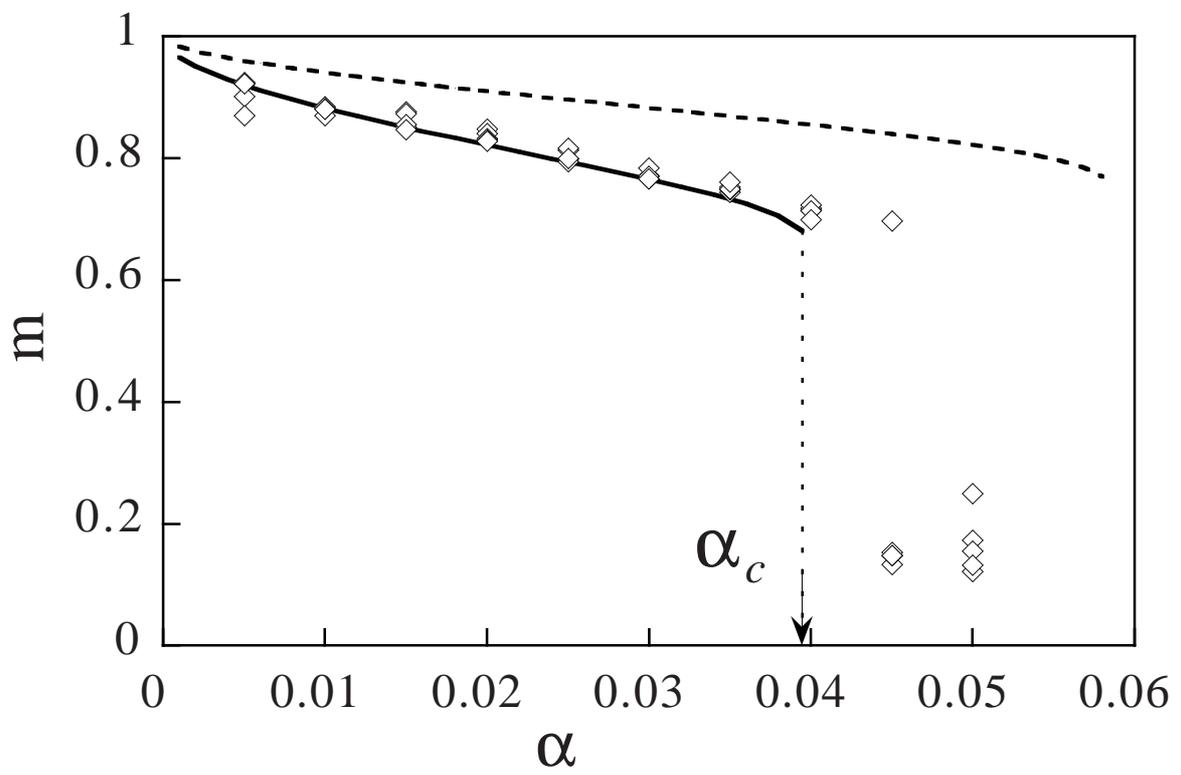
\item[FIG.2:] 
$\alpha$ dependence of the overlap $m$ obtained from the SCSNA (solid line) together
with that from numerical simulations with $N=4000$ in the case 
$p(\omega)=\delta(\omega-\omega_0)$.
To observe the result that the contribution of the $\Gamma_2$ term in 
Eq.~(\ref{local})
or (\ref{cond2}) to the value of $m$ is significant, we plot values of $m$
obtained by deliberately setting $\Gamma_2=0$ (broken line).

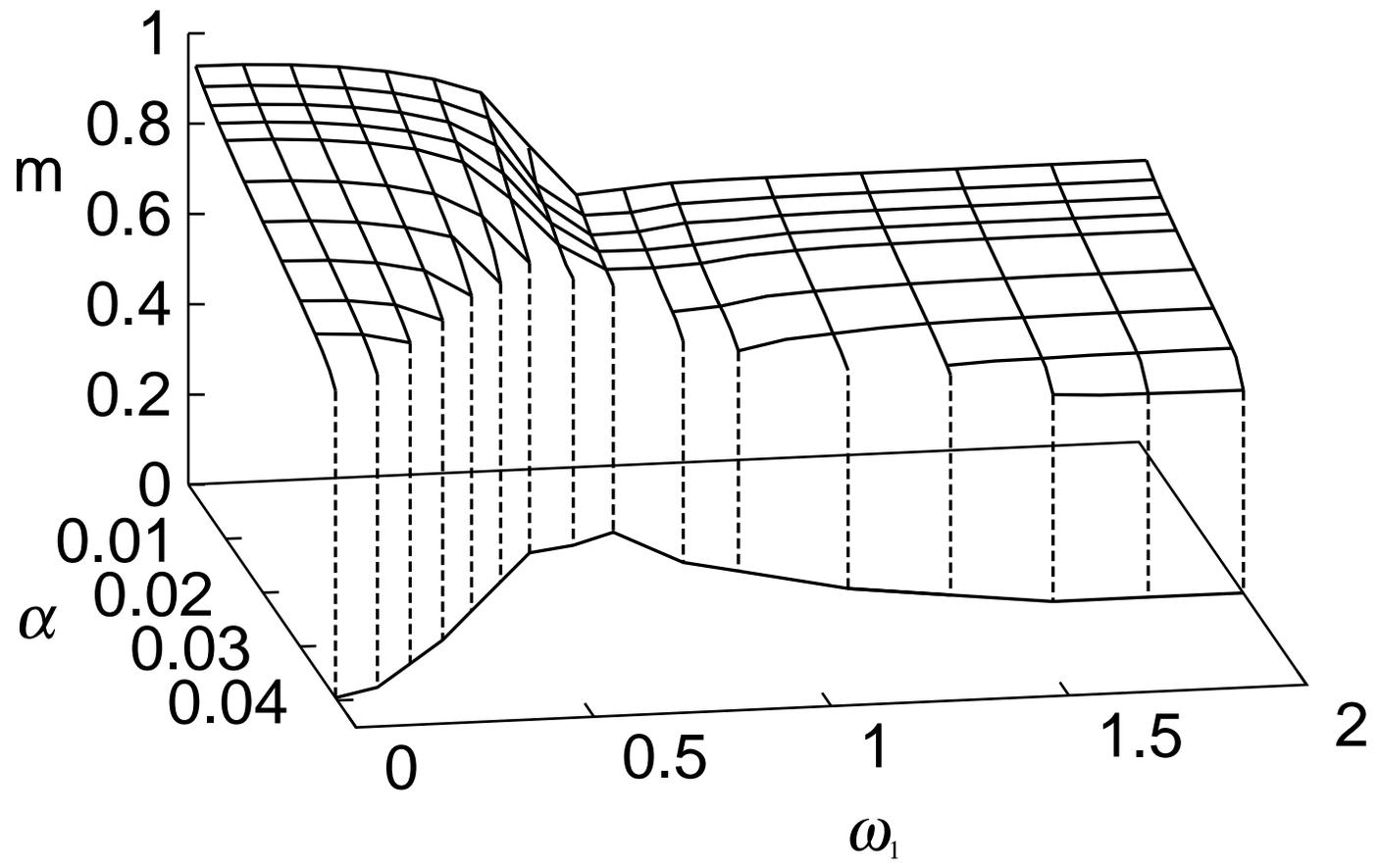
\item[FIG.3:] 
Phase diagram showing $m$-surface plotted as a function of $\alpha$ and
$\omega_1$ for the network with distributed native frequencies with
$p(\omega)=0.7\delta(\omega)+0.15\delta(\omega-\omega_1)
+0.15\delta(\omega+\omega_1)$.
We observe a valley or gap separating the small $\omega_1$
regime from the large $\omega_1$ one. 
The dependence of the storage capacities $\alpha_c$ on $\omega_1$ is represented
by the projected curve on the $\alpha$-$\omega_1$ plane.

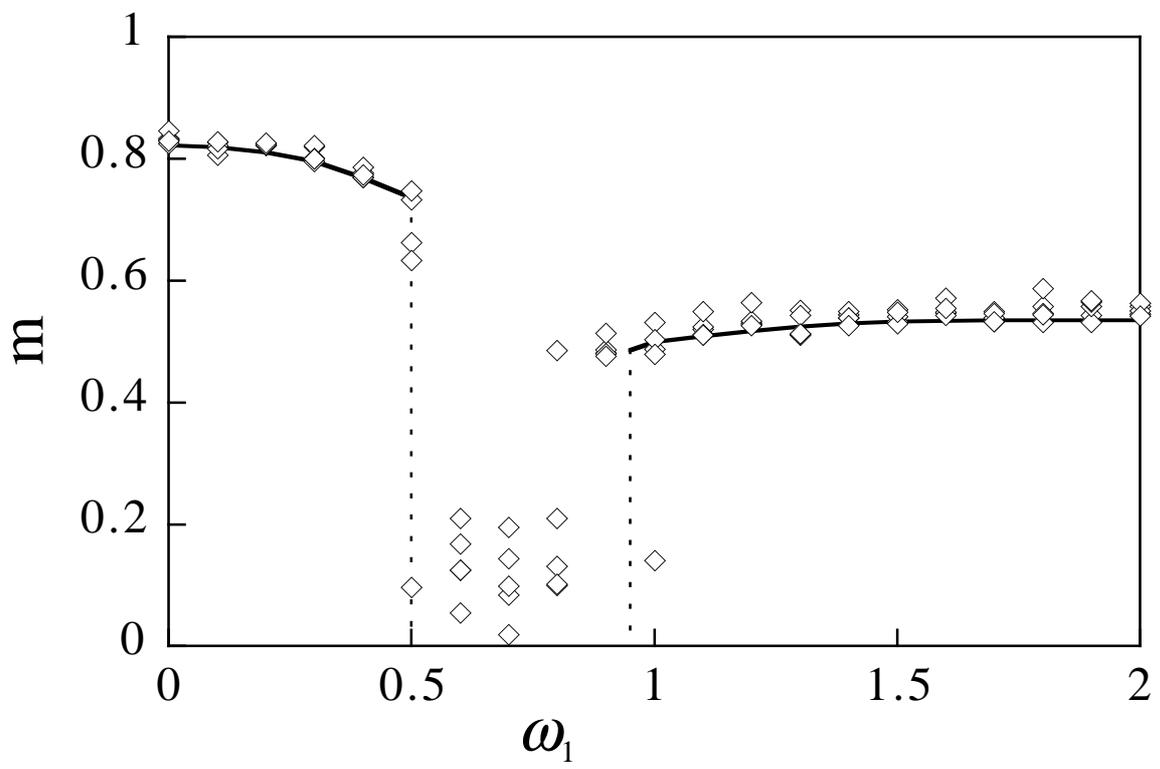
\item[FIG.4:] 
Dependence of $m$ on $\omega_1$ for the network with $\alpha=0.02$ and
$C_0=0.7$ obtained from the SCSNA (line) and numerical simulations with 
$N=2000$ (dots).
The gap separating the two regimes with
different types of synchronization are clearly depicted and is in excellent
agreement with the results of numerical simulations.

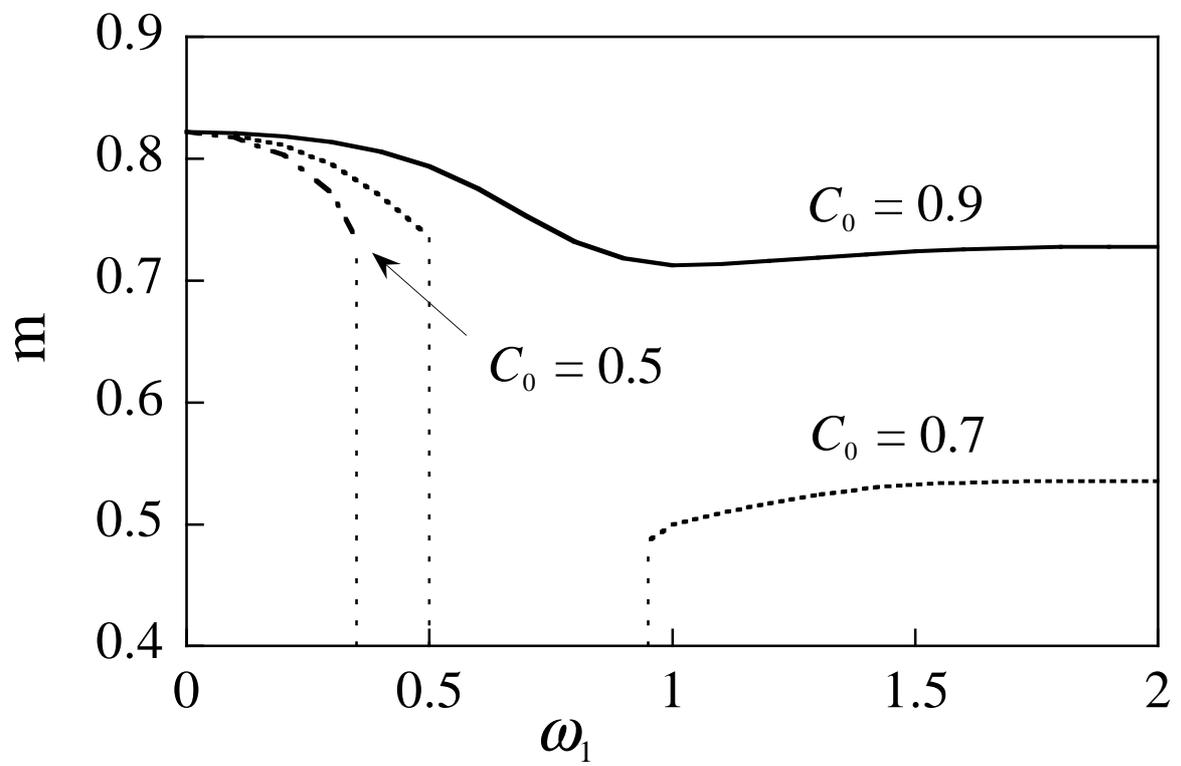
\item[FIG.5:] 
Dependence of $m$ on $\omega_1$ for $\alpha=0.02$ and $C_0=0.5, 0.7, 0.9$ obtained
from the SCSNA.

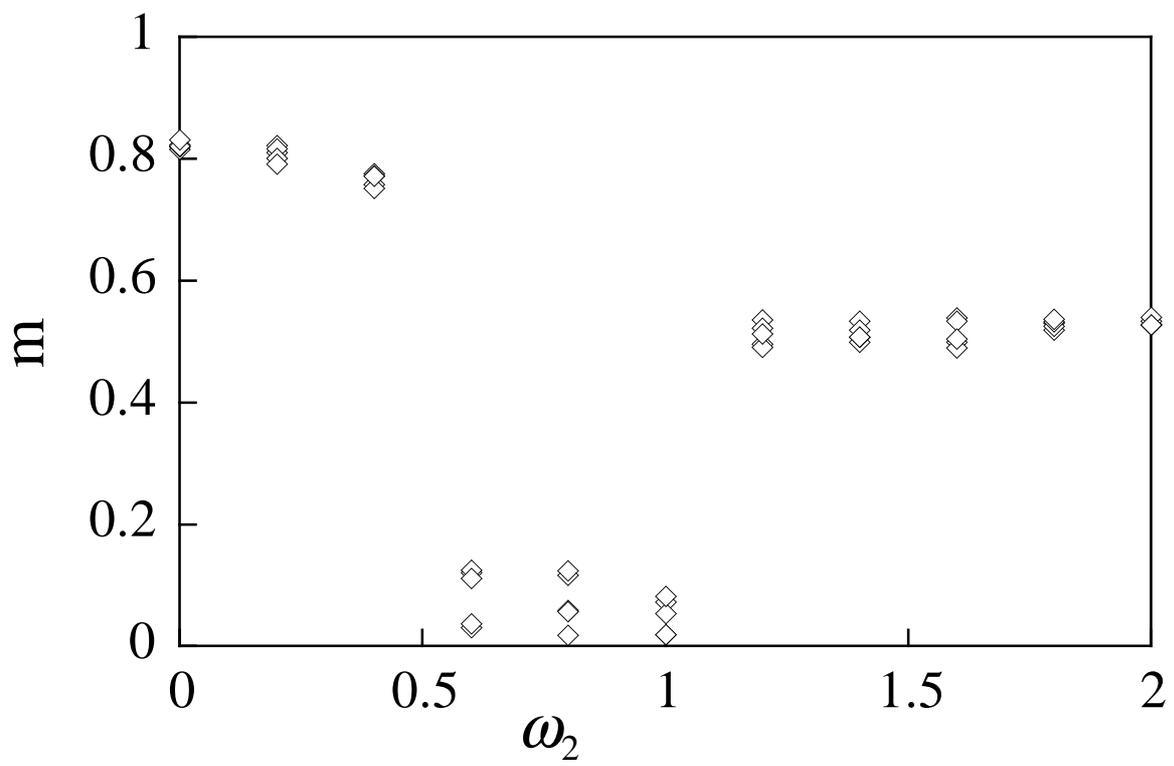
\item[FIG.6:] 
The numerical simulation result ($N=4000$) for $L=2$
showing the $\omega_2$ dependence of $m$ 
in the case with $\omega_0=0, C_0=0, \omega_1=0.1, C_1=0.35$
and $\alpha=0.02$ (see eq.~(\ref{L=2})).

\end{itemize}

\end{document}